\documentclass[%
 reprint,
superscriptaddress,
 amsmath,amssymb,
 aps,
floatfix,
]{revtex4-2}
\usepackage{xcolor}
\usepackage{graphicx}
\usepackage{dcolumn}
\usepackage{bm}

\usepackage{varioref}
\usepackage{xr-hyper}
\usepackage{hyperref}
\makeatletter

\begin{document}
\title{Effects of linking topology on the shear response of connected ring polymers:\\ Catenanes and bonded rings flow differently}

\author{Reyhaneh A.\ Farimani}
\affiliation{Department of Physics, Sharif University of Technology, Tehran, Iran}
\affiliation{Faculty of Physics, University of Vienna, Boltzmanngasse 5, 1090 Vienna, Austria}
\author{Zahra Ahmadian Dehaghani}%
\affiliation{International School for Advanced Studies (SISSA), Via Bonomea 265, 34136 Trieste, Italy}%
\author{Christos N.\ Likos}
\affiliation{Faculty of Physics, University of Vienna, Boltzmanngasse 5, 1090 Vienna, Austria}
\author{Mohammad Reza Ejtehadi}
\affiliation{Department of Physics, Sharif University of Technology, Tehran, Iran}

\date{\today}

\begin{abstract}
We perform computer simulations of mechanically linked (poly[2]catenanes, PC) and chemically bonded (bonded rings, BR)
pairs of self-avoiding ring polymers in steady shear. We find that BR's develop a novel motif, termed \textit{gradient tumbling}, rotating around the gradient axis.
For the PC's the rings are stretched and display another new pattern, termed \textit{slip-tumbling}.
The dynamics of BR's is continuous and oscillatory, whereas that of PC's is intermittent between slip-tumbling attempts. Our findings demonstrate the interplay between topology and hydrodynamics in dilute solutions of connected polymers.
\end{abstract}
\maketitle
Polymer topology has profound and fascinating implications on the equilibrium and flow properties of polymer solutions and melts~\cite{micheletti:physrep:2011,colby2003polymer}. 
A prominent example of topologically constrained polymers are (unknotted) rings, which differ from their linear counterpart through
the simple operation of joining the two ends together without any additional chemical modification or change in the solvent. 
Already this simple operation brings about spectacular changes in the properties of both the single molecule and concentrated
solutions of the same. At the dilute limit, the condition of topology conservation leads to a scaling of the ring size with molecular
weight following a self-avoiding exponent even for rings without excluded-volume interactions~\cite{deutsch:pre:1999} and it
introduces a concomitant topological potential between two rings~\cite{frank:nature:1975}.
In concentrated solutions and melts, the absence of free ends brings about once more dramatic changes in both the scaling
of the ring sizes with molecular weight~\cite{Halverson2011a} and in their conformations, which are not any more akin to 
Gaussian random walks but feature fractal, tree-like conformations \cite{Halverson2011a,Rosa2014}.
Equilibrium dynamics is affected as well, resulting in power-law stress relaxation \cite{Kapnistos2008a} 
and unique shear-thinning exponents \cite{Parisi2021}, in addition to the occurrence of mutual ring-threading events.
The latter result into the formation of reversible mechanical links 
that are responsible for 
viscosity thickening in extensional flow \cite{Huang2019,Oconnor2020} and are believed to play a key role
in the temperature-gradient vitrification in melts, resulting into the formation of an active topological glass \cite{Smrek2020a}. 
In flow-driven dilute conditions, the combination of topology with hydrodynamic interactions brings about phenomena
unique to ring polymers, such as vorticity swelling in shear, extensional or mixed flows~\cite{Hsiao2016,liebetreu:acsml:2018,young:pre:2019,tu:mm:2020} as well as hydrodynamic inflation
under steady shear~\cite{liebetreu:commats:2020}. The current interest in ring polymers is further enhanced by the biological 
relevance of circular DNA, which can be found naturally in the form of (supercoiled) bacterial plasmids \cite{Phillips2004},
extrachromosomal DNA of eukaryotes \cite{Koche2020}, or in the kinetoplast DNA \cite{klotz:pnas:2020, Chen1995a, tubiana:prx:2023} of trypanosoma. 

Going beyond simple cyclic macromolecules, mechanically interlocked~\cite{wu2017poly, Datta2020, Tranquilli2023, hart:nrm:2021, liu:csr:2022, orlandini:jpcm:2022} or chemically bonded rings~\cite{polymeropoulos:mm:2017, chen:no:2022} form a higher level of supramolecular, topological polymers. 
Advances in the synthesis of polycatenanes, macro-molecular structures made of concatenated ring polymers, have sparked  interest 
in their study, revealing 
that they exhibit distinct characteristics that differ significantly from those of traditional polymers due to their internal degrees of freedom. Dehaghanni \textit{et al.}~found that the gyration radius of polycatenane features
two distinct power-law dependencies on ring size and 
the number of rings, as a consequence of the topological
linking constraints~\cite{dehaghani:sm:2020}. In addition,  
novel types
of intramolecular, topological entanglements present in these molecules have been recently discovered~\cite{ahmadian:acsml:2023}.
Chiarantoni \textit{et al.}~studied polycatenanes in channel confinement and found that, unlike regular polymers, they continue stretching in the strong confinement 
regime~\cite{chiarantoni:mm:2023},
whereas Chen \textit{et al.}~showed 
that polycatenanes have a linear response regime to very large forces, due to delayed force penetration in these structures~\cite{chen:cjps:2023}. Moreover, polycatenane 
solutions and melts also feature intriguing dynamics and phase behavior \cite{Rauscher2020_1, rauscher2020dynamics, stano:mm:2023}. Additionally, polycatenanes are promising candidates for novel materials, such as mechanophores, molecular motors, and transmembrane ion channels \cite{zhang2020catenane, kay2008beyond, August2020}.
In recent work, Soh \textit{et al.} studied the kinetoplast under planar elongational flow,
finding that the usual abrupt coil-stretch transition in linear polymers is absent for this concatenated network~\cite{soh2020deformation}. Otherwise, very little
is known about the behavior of topologically linked,
ring-based aggregates under non-equilibrium conditions and,
in particular, under shear flow.
Here we show that for
the simplest case of two connected ring polymers, the type of linking (chemical bond vs.\ mechanical link) has 
profound consequences on the dynamical properties 
of the compound under steady shear and that novel 
dynamical patterns arise in these cases, 
unknown for any other polymer architecture. 

\begin{figure*}
    \centering
    \includegraphics{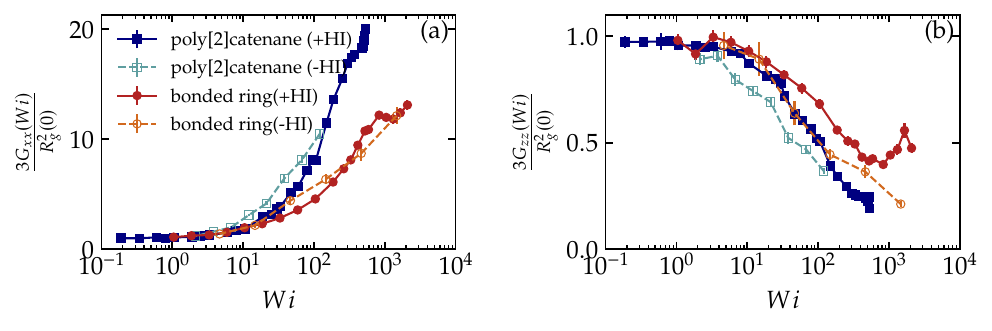}
    \caption{
    Time-averages of selected diagonal elements 
    of the gyration tensor of the poly[2]catenane and the bonded rings system (entire molecules) in shear flow,
    as a function of the Weissenberg number,
    normalized over their equilibrium values, $R_g^2(0)/3$.
    (a) The flow-direction element, $G_{xx}$;
    (b)  the vorticity direction element, $G_{zz}$.
    The dark blue-filled squares refer to 
    PC ($+{\rm HI}$), red-filled circles 
    to BR ($+{\rm HI}$), empty sky blue squares 
    to PC ($-{\rm HI}$) and empty organce circles to 
    BR ($-{\rm HI}$).}
    \label{fig:G:entire}
\end{figure*}

We consider two related
supramolecular structures: a poly[2]catenane molecule (PC), 
consisting of two linked, self-avoiding ring  polymers (Hopf link) and a system of two 
self-avoiding rings bonded by a bending- and torsion-free chemical bond (BR).
Details on the microscopic model and method are presented in the 
Supplemental Material (SM)
~\cite{[{See Supplemental Material at \href{https://drive.google.com/file/d/1FiWp7mHmclXK3g8xd0cvo9H3KNq1qzSw/view?usp=sharing}{this Link}}][{ where more details on the microscopic model, method, and relaxation time calculations can be found. Some additional parts of the results are presented there as well. See also references \cite{velocityverlet_original, Ihle2003, huang:jcp:2010, huang2015thermostat, Ripoll2005, brown:jcp:1998, rauscher:acsml:2018, ChenWenduo2017CaDo} therein.}]supp}.
We applied Multi-particle Collision Dynamics (MPCD) as well as the random solvent variant of the same
to simulate the dynamics, 
whereby the hydrodynamic interaction (HI) is included only
in the former and not in the latter~\cite{gompper2009multi, ripoll:prl:2006}. 
In what follows, the 
symbol $+{\mathrm {HI}}$ indicates the presence of HI, and $-{\mathrm {HI}}$ its absence.
Steady shear is induced by employing Lees-Edwards boundary conditions~\cite{Lees1972}, 
with $\hat{x}, \hat{y}, \hat{z}$ being
the flow, gradient, and vorticity directions, respectively.
 We also 
use the molecule's longest relaxation time
$\tau_R$ to define the dimensionless Weissenberg number 
$Wi = \tau_R \dot\gamma$.
Details on the method and the relaxation time calculation 
can also be found in the SM.

We have analyzed the
conformations and dynamics of the compounds under
steady shear at three different levels of 
description: for each ring component, 
for the entire macromolecule as well as for an effective dimer, consisting 
of the centers of mass of the two connected rings.
Information on the shape and the dynamics of any polymer can be obtained from the gyration tensor, 
whose value at time $t$ and for given $Wi$ is defined as
\begin{equation}
	G_{\alpha\beta}(t; Wi) = \frac{1}{\mathcal N} \sum_{i=1}^{\mathcal N} r^{(i)}_{\alpha}(t; Wi) 
    r^{(i)}_{\beta}(t; Wi),
\end{equation}
$\alpha,\beta \in \{x,y,z\}$, in which ${\mathcal N}$ refers to the number of monomers in the polymer, and it thus takes the value ${\mathcal N} = N$ for 
an individual ring, ${\mathcal N} = 2N$ for the whole 
compound and ${\mathcal N} = 2$ for the effective dimer. Moreover, $r^{(i)}_{\alpha}(t; Wi)$ 
denotes the 
$\alpha$ Cartesian component of the position vector 
of monomer $i$ in the polymer center of mass frame. 
The instantaneous gyration radius 
is expressed as $R_g(t; Wi) =({\rm Tr}[G(t,Wi)])^{1/2}$. 
We denote the eigenvalues of $G(t,Wi)$ with 
$\lambda_1(t; Wi) \geq \lambda_2(t; Wi) \geq \lambda_3(t; Wi)$ and 
$\hat{e}_i(t; Wi)$, $i = 1,2,3$, 
represent the corresponding 
eigenvectors.
Time averages are denoted
$G_{\alpha\beta}(Wi) = \langle G_{\alpha\beta} (t; Wi)\rangle_t$ 
and $R_g(Wi) = \langle R_g (t; Wi)\rangle_t$; at equilibrium, 
$G_{\alpha\alpha}(Wi = 0) = R_g^2(Wi = 0)/3$. We consider here
only the quantities $G_{xx}(Wi)$ and $G_{zz}(Wi)$; the results for
$G_{yy}(Wi)$ are shown in the Supplementary Figure
	S6.

Fig.~\ref{fig:G:entire}(a) reveals 
that the PC stretches along the flow axis much stronger than the BR,
illustrating
that the topology is the most important factor in enhancing stretching,
as the presence of HI does not markedly affect the amount of the stretch.
Contrary to the entire state picture,
the behavior of the individual rings comprising the PC and BR along the flow direction is practically indistinguishable, as shown in 
Supplementary Figure S4.

\begin{figure}
    \centering
\includegraphics{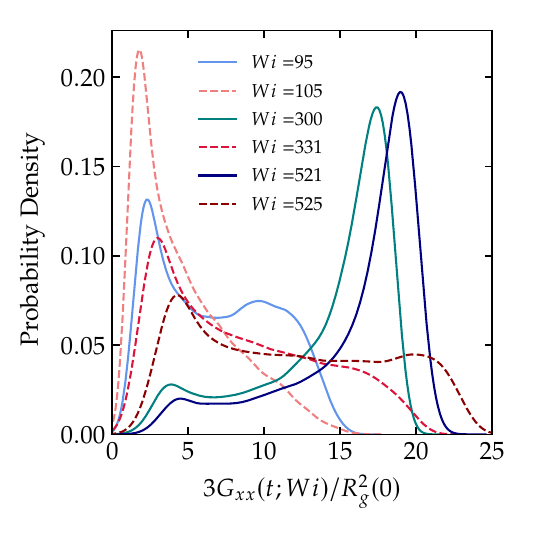}
    \caption{Probability density function of 
    $G_{xx}(t;Wi)$ normalized over $G_{xx}(0)$.
    Results are shown for PC $(+{\rm HI})$, solid lines, 
    and for BR $(+{\rm HI})$, dashed lines, at various 
    Weissenberg numbers, as 
    indicated in the legend.}
    \label{fig:Gxx:dist}
\end{figure}

The probability density function (pdf) 
of the instantaneous values 
of $G_{xx}(t;Wi)$, shown in Fig.\ \ref{fig:Gxx:dist},
reveals 
dramatic differences between the PC and the BR. Whereas the
expectation values for the PC are higher than those for the BR,
the maximal stretching of the latter exceeds that of the former.
At the same time, 
while the pdf for the PC shows a first-order-type transition 
from a state with a maximum at low values of $G_{xx}(t,Wi)$
at small $Wi$ to a state with a maximum at high values of $G_{xx}(t,Wi)$ at large $Wi$,
the maximum of the pdf's for the BR always remains at low $G_{xx}(t,Wi)$-values
and the effect of increasing $Wi$ is the stretching of the pdf to higher and higher
$G_{xx}(t,Wi)$-values with a weak secondary maximum development at the highest $Wi$-value shown.
This difference points to the presence of completely different
dynamical motifs for the two molecules under shear.

It is known that vorticity swelling~\cite{liebetreu:acsml:2018, liebetreu:commats:2020,tu:mm:2020} takes place for individual rings under shear; we found that it also occurs for the constituent rings of the BR-molecule in the $+{\rm HI}$-case but not for those of the PC, see
 Fig.S4 in the SM.
Fig.~\ref{fig:G:entire}(b) demonstrates, in addition, the presence
of vorticity swelling for
the entire molecule, again
only for the BR in the +HI-case, whereas it is
absent for the PC molecule regardless of the inclusion of HI, indicating a significant difference between the two molecules. This sets the stage for understanding a variety of dynamic features to follow. We further note that vorticity swelling is also seen in the dimer picture, as showcased in 
Supplementary Figure S5.\\
For the BR, the individual ring vorticity swelling, 
which is asynchronous for each ring due to thermal fluctuations,
initiates solvent flow on the flow-vorticity plane, which 
disrupts the other ring. 
This makes the conformation of the BR orientationally unstable and sets a rotational pattern in motion 
in which the composite molecule rotates around an axis closely aligned to the gradient
direction, extending thereby for a significant fraction
of time into the vorticity direction. It is this extension
that additionally contributes to the vorticity swelling
of the \textit{entire} bonded rings, seen 
in Fig.~\ref{fig:G:entire}(b);
a video and snapshots of this novel type of rotation can be found in the SM; see Video 1. This is a tumbling motion,
 since the entire system rotates around its center of mass \cite{degennes:jcp:1974}, but contrary to the usual tumbling
 that occurs around an axis roughly aligned with the 
 vorticity direction, this rotational motion occurs around
 an axis approximately aligned with the gradient axis. 
 Accordingly, we coin the term \textit{gradient-tumbling}
 for this dominant dynamical motif of bonded rings under
 steady shear. The usual, \textit{vorticity} tumbling,
 demonstrated in Video 2 in SM, is also
 present in the case of bonded rings, but it is not the 
 dominant dynamical pattern of motion. 
 This is
 an incessant, continuous motion, as opposed to the 
 intermittent dynamics featured by the PC.\\
The poly[2]catenane displays a very different dynamical pattern under shear, experiencing much fewer tumbling events than the bonded rings and featuring stable stretched conformations,
which result in the appearance of the second maximum
in the pdf of $G_{xx}(t; Wi)$ that dominates at high 
$Wi$-values.
The animation and snapshots from the trajectories show that the 
mechanical bond's freedom in the PC molecule plays a key role 
in allowing it to respond to the solvent-induced tension in a 
different fashion than the chemically
bonded molecule. 
In Video 3 in the SM, it can be 
seen that in the PC the two rings exchange their positions
under shear in a way we term \textit{slip-tumbling}.
Due to their particular connectivity in the form of a
Hopf-link, the two rings can undergo mutual slipping
\textit{through} each other while leaving their 
conformation tilted at a certain (low) angle with respect
to the flow direction, and they are able to maintain this
overall stretched configuration for longer intervals of
time, whereas in topologies without mechanical links, when 
the polymer alignment is opposite of the flow field, the 
polymer starts to collapse and tumble~\cite{degennes:jcp:1974}.
Occasionally, such slip-tumbling events stay incomplete
and the PC confirmation returns to the original one after
such an unsuccessful attempt, 
see Video 
4 in the SM. \\
\begin{figure*}
    \centering
    \includegraphics{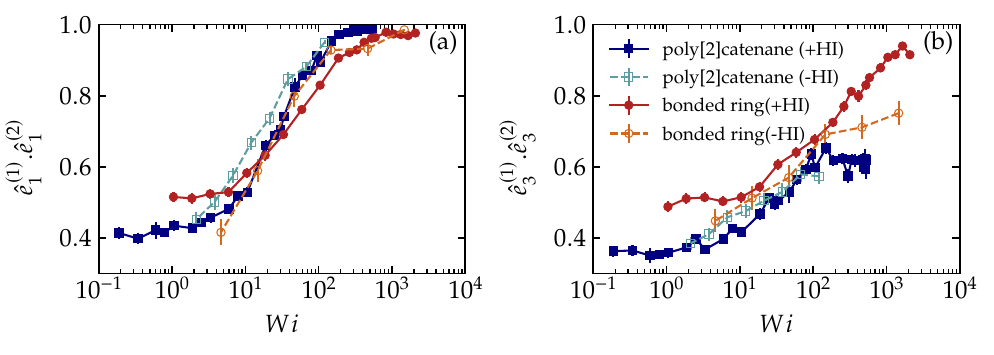}
    \caption{Time-averages $\hat{e}_{i}^{(1)} \cdot \hat{e}_{i}^{(2)}$ of the cosines 
    between the eigenvectors of the two rings
    in a PC or a BR corresponding to various
    eigenvalues. (a) The case $i = 1$, corresponding to the largest eigenvalues; 
    (b) the case $i = 3$, corresponding to the smallest eigenvalues.
    The types of molecules and the presence or absence of HI are indicated by colors and symbols
    as shown in the legend of the panel (b).}
    \label{fig:relorient}
\end{figure*}
Further evidence of the distinct response that the bonded
rings and the poly[2]catenane have to shear is offered by the
orientational correlations between the gyration eigenvectors
in the two molecules. Indeed,
vorticity swelling requires that each ring be
oriented at an angle close to the flow direction,
i.e., the eigenvectors $\hat e_{1,3}$  of
vorticity-swollen rings lie 
almost parallel to the $\hat x$ and 
$\hat y$-directions, respectively.
We consider, therefore, the time averages of the relative orientations of 
suitable eigenvectors, 
$\langle\hat{e}_{i}^{(1)}(t;Wi) \cdot \hat{e}_{i}^{(2)}(t;Wi)\rangle_t \equiv \hat{e}_{i}^{(1)} \cdot \hat{e}_{i}^{(2)}$, 
$i = 1,3$, where the 
superscripts refer to rings (1) and (2) of the
composite polymer; here we have suppressed,
for brevity, the dependence on the time-averaged 
quantity on $Wi$. Results in 
Fig.\ \ref{fig:relorient}(a) show that the 
eigenvectors corresponding 
to the largest eigenvalue become
increasingly aligned with growing $Wi$: 
starting from the value $\hat{e}_{1}^{(1)} \cdot \hat{e}_{1}^{(2)} \cong 0.5$, indicating a
random orientation at equilibrium, all curves
for both compounds and independently of
the inclusion of HI approaches 
unity
at $Wi \cong 10^3$, where the rings stretch along
the flow direction. On the other hand, 
in Fig.\ \ref{fig:relorient}(b) it can be seen that
the same does not hold true for the orientational
correlation $\hat{e}_{3}^{(1)} \cdot \hat{e}_{3}^{(2)}$ between the eigenvectors that
correspond to the lowest eigenvalue. For the BR-rings,
the curve is still monotonic and it approaches
unity at high $Wi$-values, consistently with 
the fact that they feature vorticity-swelling and thus
they resemble discs oriented almost perpendicularly to the
gradient direction. Although this joint orientation
is possible for the case of rings connected by a chemical bond, the mechanical link in the case of 
the PC renders such an arrangement highly unlikely. 
As $Wi$ grows, the mutual alignment of
the `smallest' eigenvectors grows until it reaches
a plateau-value of about 0.6 at 
$Wi \approx 100$. The presence of a Hopf link
imposes a local twist and, therefore, a strong local
penalty for the two rings
of a PC to attain disc-like shapes, both oriented
perpendicular to the $y$-axis. Instead, they
assume thin and mutually perpendicular orientations,
which suppresses therefore vorticity swelling
giving rise instead to slip-tumbling.\\
\begin{figure}
    \centering
    \includegraphics{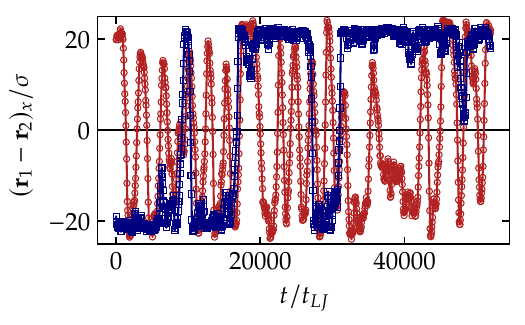}
    \caption{
    The relative $x$-component (flow direction) of the rings' centers of mass as a function of time for $Wi \approx 500$. The red line corresponds to BR $(+{\rm HI})$ at $\dot\gamma = 10^{-1.3}$
    ($Wi = 525$)
    and the blue one to PC $(+{\rm HI})$
    at $\dot\gamma = 10^{-1.06}$
    ($Wi = 521$).
    }
    \label{fig:tumbling:event}
\end{figure}
Strong evidence for the different types of dynamics 
featured by bonded rings and poly[2]catenanes is offered
by looking at representations of typical trajectories,
i.e., time series of certain characteristic quantities of the 
motion. 
For this purpose, the effective dimer representation is the most suitable, as it employs a minimalistic description that captures nevertheless the salient 
properties of the system. First, we consider the 
$x$-component of the relative separation 
${\mathbf r}_1 - {\mathbf r}_2$ between the two poles
of the dimer, shown as a time series for $Wi \cong 500$
in Fig.\ \ref{fig:tumbling:event}. 
The PC features two stable configurations with 
occasional flipping events caused by slip-tumbling and thus
it is characterized by intermittent dynamics, whereas the 
BR features incessant rotational motion characteristic of
the dominant, gradient-tumbling pattern of the same. 
We further consider the alignment angle $\theta(t;Wi)$ of the 
composite molecules with the flow axis by employing
the instantaneous version of the time-average 
relation \cite{teixeira:mm:2005}:
\begin{equation}
    \theta(t;Wi) = \frac{1}{2}
    \tan^{-1}\left[\frac{G_{xy}(t;Wi)}
    {G_{xx}(t;Wi) - G_{yy}(t;Wi)}\right].
\end{equation}
Results at $Wi \approx 500$ are presented in 
Fig.\ S7 in SM. The alignment angle 
of the PC is not only smaller in magnitude than that
of the BR, but it features much less pronounced fluctuations
as well, affirming the property 
that the PC has a much
more stable orientation in space than the BR, a feature
arising from its unique linking topology, allowing it
to slip-tumble and thus avoid turbulent tumbling events
\cite{degennes:jcp:1974,schroeder:prl:2005,teixeira:mm:2005}. 
In addition, the angle is essentially always positive for
both cases, indicating that vorticity-tumbling is extremely rare.
A quantitative analysis of the characteristic tumbling frequencies is presented in Supplementary Figure
S8.

A reduced frequency of slip-tumbling events and 
suppressed shape fluctuations during slipping (PC), as opposed to incessant gradient- and occasional vorticity-tumbling
(BR) explain the dramatic differences
in the pdf's of the two, shown in Fig.~\ref{fig:Gxx:dist}.
This observation is even more intriguing if compared to the results by Chen \textit{et al.}~\cite{chen:cjps:2023}, who recently studied poly[n]catenane and bonded rings systems under extensional forces \textit{in equilibrium}.
It was found that the elastic modulus of poly[2]catenane is nearly double that of the bonded rings, i.e.,
for a given external force, the BR stretches more than the PC.
In our case, for given $Wi$, the BR can 
achieve longer maximum stretching compared to the PC,
as seen in Fig.~\ref{fig:Gxx:dist}.
However, the time average shows the opposite behavior due to tumbling events, a hydrodynamically-induced pattern 
of motion absent when
simply pulling the macromolecule by a bare force at both ends, as is the case in the work of 
Ref.~\cite{chen:cjps:2023}.

We have carried out computer simulations of chemically bonded and mechanically linked ring polymers under shear with full consideration of the hydrodynamic interactions, establishing that connected polymer dynamics is very rich and 
strongly specific on the linking architecture. 
These structures experience different pathways of tumbling 
unknown to other polymer topologies.
Mechanically linked systems feature a stable stretched state 
and are independent of the presence or absence of hydrodynamic 
interactions and feature slip tumbling. Accordingly, we 
expect systems that possess a high number of mechanical bonds 
not to experience tumbling under shear flow,
consistently with the observed behavior of kinetoplast under 
elongational flow~\cite{soh2020deformation}. 
While both chemically bonded and mechanically linked systems exhibit shear-thinning behavior,
the latter maintain their stretched state for longer time intervals, resulting in a higher shear 
viscosity than their chemically bonded counterparts.  
On the other hand, chemically bonded ring pairs are highly 
agile under flow but, at odds with usual polymer dynamics,
they display gradient-direction tumbling.
Based on the fluctuating origin of this type of tumbling, we 
expect that bonding more and more rings chemically in a linear 
fashion would eventually suppress the emergence of such 
dynamics. 
Our results underline the intriguing interplay between
topology, hydrodynamics, and thermal fluctuations in
polymer solutions, and they are, in principle 
experimentally observable in suitably constructed 
microfluidic devices~\cite{tu:mm:2020, tanyeri:nl:2013}.

 We acknowledge support from the European Union (Horizon-MSCA-Doctoral  Networks) through the project QLUSTER  (HORIZON-MSCA-2021-DN-01-GA101072964).
The calculations were performed at the Vienna Bio Center Cluster and the Vienna Scientific Cluster (VSC). 
M.R.E.~acknowledges
support from the ICTP (Trieste) through the Associates Programme (2019-2025).   
\bibliography{main}

\end{document}